\documentclass[aps,twocolumn,pre,showpacs,groupedaddress,showkeys,amsmath,amssymb,superscriptaddress]{revtex4}
\usepackage{graphicx}
\usepackage{bm}
\usepackage{epic}
\usepackage{eepic}
\usepackage{pifont}
\usepackage[latin1]{inputenc}
\usepackage{rotating}
\usepackage{color}
\usepackage{nicefrac}
\usepackage{ulem}
\usepackage{subfig}
\input epsf
\input rotate
\usepackage{graphicx}

\begin{document}

\title{How does confinement affect the dynamics of viscous vesicles and red blood cells?}

\author{Badr Kaoui}
\email{b.kaoui@tue.nl}
\affiliation{Department of Applied Physics, Eindhoven University of Technology, P.O. Box 513, 5600 MB Eindhoven, The Netherlands}
\author{Timm Kr\"{u}ger}
\email{t.kruger@tue.nl}
\affiliation{Department of Applied Physics, Eindhoven University of Technology, P.O. Box 513, 5600 MB Eindhoven, The Netherlands}
\author{Jens Harting}
\email{j.harting@tue.nl}
\affiliation{Department of Applied Physics, Eindhoven University of Technology, P.O. Box 513, 5600 MB Eindhoven, The Netherlands}
\affiliation{Institute for Computational Physics, University of Stuttgart, Pfaffenwaldring 27, D-70569 Stuttgart, Germany}
\date{\today}
\begin{abstract}
Despite its significance in microfluidics, the effect of confinement on the transition from the tank-treading (steady motion) to the tumbling (unsteady motion) dynamical state of deformable micro-particles has not been studied in detail. In this paper, we investigate the dynamics of a single viscous vesicle under confining shear as a general model system for red blood cells, capsules, or viscous droplets. The transition from tank-treading to tumbling motion can be triggered by the ratio between internal and external fluid viscosities. Here, we show that the transition can be induced solely by reducing the confinement, keeping the viscosity contrast constant. The observed dynamics results from the variation of the relative importance of viscous-, pressure-, and lubrication-induced torques exerted upon the vesicle. Our findings are of interest for designing future experiments or microfluidic devices: the possibility to trigger the tumbling-to-tank-treading transition either by geometry or viscosity contrast alone opens attractive possibilities for microrheological measurements as well as the detection and diagnosis of diseased red blood cells in confined flow.
\end{abstract}

\maketitle
\section{Introduction}
The dynamics of deformable micro-particles (e.g., droplets, polymeric capsules, phospholipid vesicles, or blood cells) under flow is a fascinating fundamental problem with increasing relevance for technological applications, for example, in food processing, drug delivery, or designing lab-on-chip devices. The present paper constitutes a contribution to the study of vesicle and red blood cell (RBC) dynamics, an attractive subject that has led to numerous publications in the last two decades (see \cite{Vlahovska2009} and the references therein). A key aim of these works is to derive constitutive physical laws to bridge the details of the dynamics of each single particle (at the microscale) to the global behavior of a suspension (at the macroscale). This bottom-up approach is generally used to study the dynamics of so-called complex fluids (e.g., blood) since the classical way of treating these as a single continuum medium fails to capture the relevant physics. It is required especially when the fluid is confined to length scales comparable to the particle size, for instance, blood flow in the microcirculatory system. The same scenario is faced in lab-on-chip devices where the composed fluid (carrying fluid with suspended particles) flows in micro-channels with cross sections of similar size as that of the transported particles. Moreover, it was found that confinement - due to the presence of the walls - alters the dynamics (see \cite{Jendrejack2003,Peyla2007} for polymers) and stability of deformable micro-particles (see \cite{Sibillo2006,Janssen2008,Janssen2010} for droplets).

The present study is devoted to investigate the interplay between the confinement and the viscosity contrast (ratio between internal and external fluid viscosities) on the dynamics of viscous vesicles and RBCs. This study is relevant to elucidate, for example, the challenge of the reproducibility and the accuracy of blood viscosity measurements that results from the interplay between rheometer wall effects and the dynamical state of each individual RBC. On the one hand, confinement can lead to an overestimation of the measured effective viscosity of blood as mentioned in \cite{Forsyth2011}. On the other hand, the transition of the dynamical states of vesicles and RBCs induced by varying the viscosity contrast was found to minimize the effective viscosity \cite{Vitkova2008}. However, to our knowledge, a systematic study that takes into account both effects (confinement and viscosity contrast) is lacking. It is known \cite{Keller1982,Beaucourt2004,Mader2006,Kantsler2006} that an unconfined vesicle or RBC subjected to shear flow undergoes either a steady liquid-like motion called \textit{tank-treading} (the vesicle main axis assumes a steady inclination angle with the flow direction while the membrane undergoes a tank-treading-like motion) or unsteady solid-like motion called \textit{tumbling} (the vesicle flips as a solid particle \cite{Jeffery1922}). These dynamical states depend on three dimensionless control parameters: (i) the swelling degree $\Delta$, (ii) the viscosity contrast $\Lambda$, and (iii) the capillary number ${\rm Ca}$ (ratio between viscous and membrane bending forces) \cite{Noguchi2007,Lebedev2007,Kaoui2009,Deschamp2009}. The transition between the two dynamical states can be induced by varying these three control parameters. In this paper, we show that there is an additional control parameter: the degree of confinement $\chi$.

In the reported simulations, we place a vesicle in a linear shear flow. This way, one avoids the interplay between the vesicle deformability and nonlinear velocity profiles which would lead to complex dynamics, even in the absence of a viscosity contrast \cite{Kaoui2011b,Shi2012}. We keep the Reynolds number, the capillary number, the shear rate, and the swelling degree unchanged. Consequently, we examine how the system reacts to the variation of only two control parameters: the viscosity of the enclosed fluid (an intrinsic vesicle property) and the degree of confinement (ratio of the vesicle size to the height of the channel). The effect of confinement on the transition of vesicle dynamics was briefly studied by Beaucourt \textit{et al}. \cite{Beaucourt2004} using phase-field simulations: For given $\chi$, the transition was induced by varying $\Lambda$. In contrast, we induce the transition by varying only the confinement, keeping all fluid and vesicle properties (especially $\Lambda$) unchanged.  This provides the interesting possibility to trigger the transition in experiments by merely adjusting the wall distance without replacing the vesicle or the suspending fluid. Additionally, we report how the physical quantities characterizing each dynamical state vary with confinement.
\section{Model and method}
Vesicles are closed membranes made of phospholipid molecules, e.g., dioleoylphosphatidylcholine (DOPC). They constitute a good biomimetic model for living cells, for example, to study blood cell dynamics. They are also used for micro-encapsulation of active materials in drug delivery. At room temperature ($\sim 25^\circ \text{C}$), most of the phospholipid membranes, in particular biological membranes, are in the liquid state. Their thickness ($\sim 5\, \text{nm}$) is negligibly small compared to the vesicle size (typically $\sim 10\, \mu \text{m}$ for giant unilamellar vesicles). Therefore, vesicle membranes are considered as two-dimensional incompressible Newtonian liquids. As a consequence, the membrane area cannot undergo extension or compression since such deformation modes would cost a lot of energy. This implies local and global conservation of the vesicle area, as is the case for RBCs \cite{Fischer1992}. Mathematically and numerically, this constraint can be achieved by using a local Lagrangian multiplier $\sigma$ that plays the role of an effective surface tension. The associated tension energy is given by $E_S=\int_{\partial \Omega} \sigma (s)\, \text{d}s$, where the integration is performed over the membrane surface $\partial \Omega$. The mechanical deformation mode that costs less energy is bending, $E_B=\frac{\kappa _B}{2} \int_{\partial \Omega} c^2 (s)\, \text{d}s$, where $\kappa _B$ is the bending rigidity and $c(s)$ the local membrane curvature. When the membrane is bent due to hydrodynamic stresses, and therefore brought out of its equilibrium configuration, it exerts a reaction force back on its surrounding fluid. The force is derived by taking the functional derivative $\textbf{F}={\delta (E_B + E_S)}/{\delta \textbf{r}}$. In two-dimensions (2D) this results in \cite{Kaoui2011a}
\begin{equation}
\mathbf{F} \!=\! \left[\! \kappa _B \!\left(\frac{\partial^2 c}{\partial s^2} +
\frac{{c}^3}{2}\right) \!-\! c \sigma \right]\!\mathbf{n} +
\frac{\partial\sigma}{\partial s}\mathbf{t},\!
\label{eq:force}
\end{equation}
where $\mathbf{n}$ and $\mathbf{t}$ are, respectively, the unit normal and tangent vectors. Our present model can be applied to vesicles as well as to RBCs. In 2D, both particles exhibit similar dynamics \cite{Kaoui2011a,Kaoui2011b,Shi2012}. However, in three-dimensional models, we need to account for the RBC's shear elasticity. This leads, in some situations, to dynamics not observed for vesicles (with zero shear elasticity), for example, the \textit{swinging} motion \cite{Forsyth2011}. 

The vesicle membrane is a free-moving boundary whose shape and position is not known a priori. Its dynamics results from its interaction with the flow of the surrounding fluid. In the present work, the membrane dynamics is computed using a front-tracking method in 2D. The membrane is represented by a moving Lagrangian mesh (a contour in 2D) and the fluid flow is computed on a fixed Eulerian lattice. Instead of solving the Navier-Stokes equations inside and outside the vesicle directly, we use the lattice-Boltzmann method (LBM). The LBM is commonly employed to study the dynamics and rheology of complex fluids (see, e.g., \cite{Aidun2010,Krueger2011,Frijters2012}). Both the position and velocity spaces are discretized; here we use nine discrete velocity directions in 2D. The main quantity in the LBM is the density distribution $f_i(\textbf{r},t)$ giving the probability to find an elementary portion of the fluid at position $\textbf{r}$ with velocity in direction $\textbf{e}_i$. The time evolution of $f_i$ is governed by the lattice-Boltzmann equation with the Bhatnagar-Gross-Krook (BGK) collision operator \cite{Succi2001}
\begin{equation}
f_{i}(\textbf{r}+\textbf{e}_i \Delta t ,t + \Delta t)=f_i(\textbf{r},t)-\frac{\Delta t}{\tau}\left[f_i(\textbf{r},t) - f_i^{\text{eq}}(\textbf{r},t)\right],
\label{eq:boltzmann}
\end{equation}
where $\Delta t$ is the time step. $\tau$ is a relaxation time related to the dynamic viscosity $\eta$ via the relation $\eta = \rho c_s^2(\tau - 1/2)\Delta x ^2/\Delta t$, where $\rho=\sum_{i=0}^{8}f_{i}$ is the fluid mass density, $c_s$ is the speed of sound and $\Delta x$ the grid spacing. $f_{i}^{\text{eq}}$ is an equilibrium distribution. The local fluid velocity $\textbf{u}=\frac{1}{\rho}\sum_{i=0}^{8}f_i\textbf{e}_i$. Geometrical quantities of the membrane (e.g. the local curvature $c$) required to evaluate the force given by Eq.~(\ref{eq:force}) are computed using the finite difference method (FDM). The membrane force $\textbf{F}$ acting on the fluid is included in the model by adding the term ($\textbf{F} \cdot \textbf{e}_i$) to the right-hand-side of Eq.~(\ref{eq:boltzmann}). The two-way coupling between the fluid flow and the membrane dynamics is accomplished using Peskin's immersed boundary method (IBM) \cite{Peskin1977}. The physical quantities computed in each mesh are matched by interpolation. More details about the numerical methods can be found in Ref.~{\cite{Kaoui2011a}}. In the present work, in contrast to our previous article \cite{Kaoui2011a}, we consider that the viscosities of the internal fluid $\eta_{\rm int}$ and of the external fluid  $\eta_{\rm ext}$ are different ($\Lambda = \eta_{\rm int} / \eta_{\rm ext} \neq 1$). We locate the fluid nodes relative to the vesicle (inside or outside) by the \textit{even-odd rule} \cite{ORourke1998}. Then the LBM relaxation time is set depending on the current fluid node location. Unlike in phase-field or level-set methods, the even-odd algorithm does not rely on diffuse scalar fields for the viscosity, and no additional field equation has to be solved. To our knowledge, this is the first time the dynamics of viscous vesicles is studied using a combination of the lattice-Boltzmann and immersed boundary methods. 

In the following, a single neutrally buoyant vesicle is positioned at mid-distance between two parallel plates. A linear shear flow with desired shear rate $\gamma$ is generated by moving the two plates in opposite directions. All the results shown below have been obtained for a Reynolds number of ${\rm Re}=\rho \gamma R_0^2 /\eta _{\rm ext} = 0.05$ (negligible inertia) and a capillary number of ${\rm Ca}= \eta_{\rm ext} \gamma R_0^{3}/\kappa _B=0.5$ (low deformation regime), where $R_0$ is the effective vesicle radius. In 2D, $R_0=P/2\pi$, where $P$ is the vesicle perimeter. In the low deformation regime, the vesicle shape is always close to its equilibrium configuration and the deformation does not alter the dynamics. The swelling degree $\Delta = 4\pi A/P^2$ ($A$ is the vesicle area) is chosen as $\Delta = 0.8$ in order to compare with the data in \cite{Beaucourt2004}, which to the best of our knowledge is the only available paper where the dynamics of a confined viscous vesicle under shear flow is studied. The external viscosity $\eta_{\rm ext}$ is also fixed in all simulations. In this paper, we vary only the two key controlling parameters: (i) the viscosity contrast $\Lambda$ and (i) the degree of confinement $\chi$. The viscosity contrast $\Lambda$ is set to values between $0.5$ and $30$ by varying only the internal viscosity $\eta_{\rm int}$. For healthy RBCs under physiological conditions, $7 \leq \Lambda \leq 13$. The confinement is defined as $\chi = R_0/W$, where $W$ is the channel half-height. 

\section{Results and discussion}
\begin{figure}[t]
\centerline{\includegraphics[width=.28\textwidth,angle=-90]{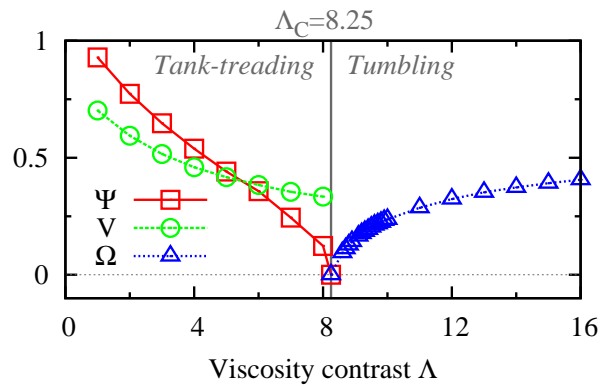}}
\caption{\label{fig:bifurcation_visco}
The transition of the vesicle dynamical state induced by increasing the viscosity contrast $\Lambda$ at fixed confinement $\chi =0.26$. Left panel: Both the inclination angle $\Psi$ and the tank-treading velocity $V$ decrease with $\Lambda$. Right panel: The tumbling frequency $\Omega$ increases with $\Lambda$. Qualitatively, the same dynamical behavior is observed for unconfined viscous vesicles ($\chi =0$) \cite{Keller1982,Beaucourt2004,Mader2006,Kantsler2006}. All plotted quantities are normalized (see text for details).}
\end{figure}

First, we investigate the effect of the viscosity contrast $\Lambda$. We compute the physical quantities characterizing each vesicle dynamical state when varying $\Lambda$, while keeping the confinement fixed at $\chi=0.26$. The tank-treading motion is characterized by the steady inclination angle $\Psi$ and the membrane tank-treading velocity $V$, while the tumbling motion is characterized by the tumbling frequency $\Omega$. Fig.~\ref{fig:bifurcation_visco} shows the variation of these quantities when increasing $\Lambda$ from $1$ to $16$. For convenience, $\Psi$ is normalized by $\pi/6$, $V$ and $\Omega$ are normalized, respectively, by the rotational velocity ($\gamma R_0/2$) and frequency ($\gamma/4\pi$) of a rigid cylinder rotating in unbounded shear flow \cite{Cox1968}. At lower $\Lambda$, the vesicle performs the tank-treading motion (left panel). By increasing $\Lambda$, both $\Psi$ and $V$ decrease. $\Psi$ decreases until it vanishes at a critical value $\Lambda_C$ (here, $\Lambda _C = 8.25$) above which tumbling takes over (right panel). In the tumbling regime, $\Omega$ increases with $\Lambda$ until saturation at larger $\Lambda$ (limit of a rigid particle). Even at this finite degree of confinement ($\chi=0.26$), we are able to reproduce qualitatively the same dynamical behavior known for unconfined viscous vesicles ($\chi =0$), reported for example in Refs.~\cite{Keller1982,Beaucourt2004,Mader2006,Kantsler2006}. 

\begin{figure}[b]
\resizebox{\columnwidth}{!}{\includegraphics[angle=-90]{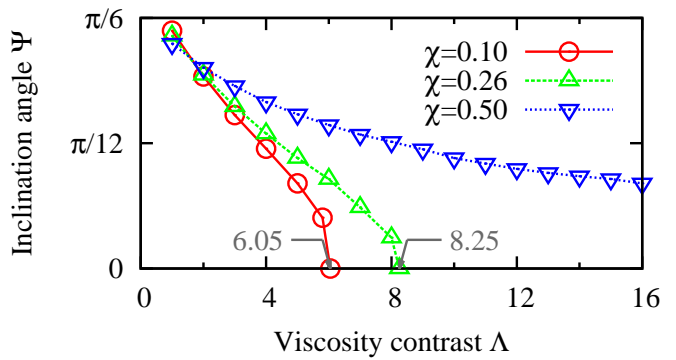}}
\caption{\label{fig:angle_visco}
The inclination angle $\Psi$ (in radians) versus the viscosity contrast $\Lambda$ of tank-treading vesicles at three different degrees of confinement ($\chi = 0.10$, $0.26$ and $0.50$). The transition threshold (where $\Psi$ vanishes) shifts to higher values when increasing confinement.}
\end{figure}

Next, we examine the effect of confinement. We compute $\Psi$ of tank-treading vesicles as a function of $\Lambda$ and at three different values of $\chi$ ($0.10$, $0.26$, $0.50$). Fig.~\ref{fig:angle_visco} shows the decreasing trend of $\Psi$ with increasing $\Lambda$ for all three degrees of confinement, as seen in Fig.~\ref{fig:bifurcation_visco} for $\chi=0.26$. However, we notice that for both $\chi=0.10$ and $0.26$, the inclination angle vanishes at two different values of $\Lambda$. By increasing $\chi$ from $0.10$ to $0.26$, the threshold of the tank-treading to tumbling transition shifts from $6.05$ to $8.25$. For the larger value $\chi=0.5$, $\Psi$ decreases as well but without reaching zero, therefore, the transition to tumbling motion is not observed. From Fig.~\ref{fig:angle_visco}, we learn that confinement increases $\Lambda _C$ and even eliminates tumbling at higher confinement (for the parameter range explored in the present study).

\begin{figure}[t]
\resizebox{\columnwidth}{!}{\includegraphics[angle=-90]{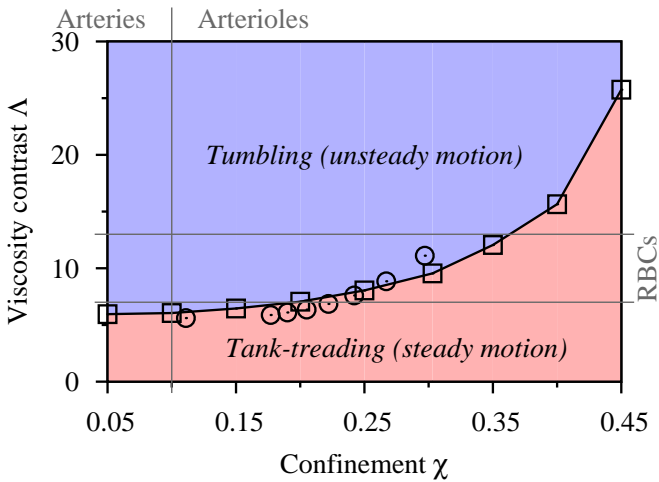}}
\caption{\label{fig:transition_conf}
The $\chi$-$\Lambda$ phase-diagram of the dynamical states (tank-treading or tumbling) of confined viscous vesicles subjected to shear flow. The confinement shifts or even inhibits the transition to tumbling. The range of $7 \leq \Lambda \leq 13$ corresponds to healthy RBCs. Typical confinement in arteries and arterioles are also shown. Circle symbols are data from Ref.~\cite{Beaucourt2004}.}
\end{figure}

In Fig.~{\ref{fig:transition_conf}}, the critical viscosity contrast $\Lambda_C$ is plotted against the confinement $\chi$. This gives the $\chi$-$\Lambda$ phase-diagram of the dynamical states of a confined viscous vesicle subjected to shear flow. It can be seen that the transition threshold $\Lambda_C$ is pushed up by increasing confinement. The same trend has been observed by Beaucourt \textit{et al} \cite{Beaucourt2004} for $0.1 \leq \chi \leq 0.3$. A comparison of the present data obtained, by IBM/LBM (square symbols in Fig.~{\ref{fig:transition_conf}}), with those reported in \cite{Beaucourt2004} by means of the phase-field method (circle symbols in Fig.~{\ref{fig:transition_conf}}) reveals a good qualitative and quantitative agreement.  We consider a more extended interval of confinement $0.05 \leq \chi \leq 1.25$ as compared to $0.1 \leq \chi \leq 0.3$ reported in \cite{Beaucourt2004}. Fig.~{\ref{fig:transition_conf}} leads to the interesting finding that the transition between the tank-treading and the tumbling motion can also be induced by varying the confinement $\chi$ alone, without changing the viscosity contrast $\Lambda$. For example, by taking a tumbling vesicle with $\Lambda > \Lambda _C$ at lower confinement, we can force it to tank-tread by increasing the confinement above a certain threshold $\chi _C$. We expect the same confinement-induced transition (beside the shear-induced transition) to happen for RBCs, for which $7 \leq \Lambda \leq 13$, during their displacement from arteries (low confinement: $2W > 100 {\rm \mu m}$, $\chi < 0.1$) to arterioles (high confinement: $10 {\rm \mu m}  < 2W < 100 {\rm \mu m}$, $ 0.1 < \chi < 1$).

\begin{figure}[t]
\resizebox{\columnwidth}{!}{\includegraphics[angle=-90]{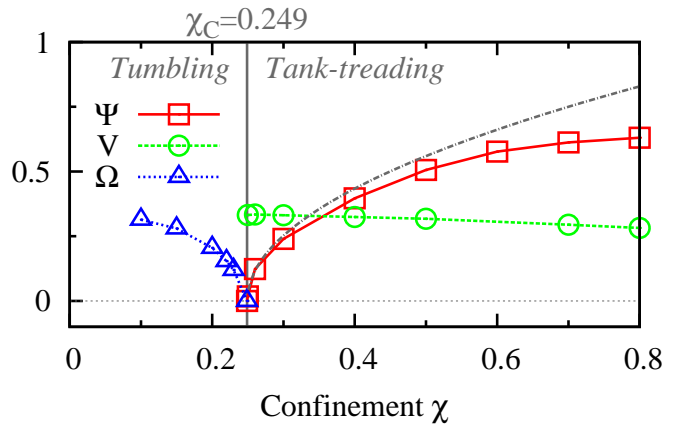}}
\caption{\label{fig:bifurcation_conf}
The transition from tumbling to tank-treading of a viscous vesicle ($\Lambda =8$) induced by confinement $\chi$ (in contrast to Fig.~\ref{fig:bifurcation_visco} where the transition is induced by varying $\Lambda$). Left panel: The tumbling frequency $\Omega$ decreases down to zero with confinement. Right panel: The angle $\Psi$ increases while the tank-treading velocity $V$ decreases with confinement. The dashed-dotted
line is fit of the angle with a square root law $\sim \sqrt{\chi - \chi_C}$. All plotted quantities are normalized (see text for details).}
\end{figure}

The transition due to confinement occurs via a saddle-node bifurcation as shown in Fig.~\ref{fig:bifurcation_conf}, for a vesicle with $\Lambda=8$. The angle $\Psi$ is well fitted with a square root law $\sim \sqrt{\chi - \chi _{\rm C}}$ in the vicinity of the bifurcation point $\chi _{\rm C}$. At lower confinement (left panel), the vesicle tumbles. By increasing $\chi$, the vesicle motion starts to be affected by the presence of the walls. The tumbling frequency $\Omega$ (triangle symbols with dotted line) decreases until it vanishes at the transition point ($\chi_{\rm C} = 0.249$). Beyond this threshold, and at higher confinement (right panel), the tank-treading motion takes over. It has to be stressed that this was not expected to happen at $\chi=0.249$. At this degree of confinement, the gap between the membrane and the wall is three times the vesicle size $R_0$. This available free space would be sufficient for a full tumbling period of the vesicle. A physical contact between the vesicle and the walls is possible only for $\chi \geq \chi ^{*}$, where $\chi ^{*}=0.75$ is the confinement for which the vesicle long semi-axis and the channel half-height $W$ are equal. Thus, we conclude that the tumbling-to-tank-treading transition induced by confinement is not entirely due to geometrical constraints. 
Fig.~\ref{fig:bifurcation_conf} (right panel) reveals another astonishing observation. The inclination angle $\Psi$ is found to increase with confinement. This result is in contradiction with our previous work for non-viscous vesicles ($\Lambda=1$) where the confinement is found to reduce the angle rather than increasing it \cite{Kaoui2011a}. Furthermore, in Fig.~\ref{fig:bifurcation_conf}, the tank-treading velociy $V$ evolves in the opposite way. It decreases with confinement. However, this trend is consistent with the results in \cite{Kaoui2011a}. The intriguing behavior in Fig.~\ref{fig:bifurcation_conf} is the opposite variation in $\Psi$ and $V$ with confinement $\chi$, unlike in Fig.~\ref{fig:bifurcation_visco} where both quantities decrease with the viscosity contrast $\Lambda$. In Ref.~{\cite{Kaoui2011a}}, the decrease of the inclination angle with confinement was explained to be caused by a torque exerted by the walls upon the vesicle. Obviously, this explanation does not hold for viscous vesicles ($\Lambda > 1$). The arising question is then: How does the torque incline the vesicle with the flow for $\Lambda=1$ while it does the opposite for $\Lambda>1$? In both situations, the walls, the external fluids, and the membrane properties are identical. The only difference is the internal viscosity $\eta _{\rm int}$ and the resulting viscosity contrast $\Lambda$. 
\begin{figure}
\resizebox{\columnwidth}{!}{\includegraphics[angle=-90]{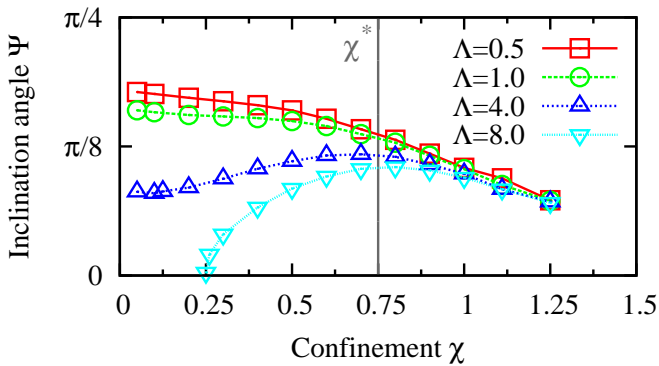}}
\caption{\label{fig:angle_conf}
The variation of the inclination angle $\Psi$ (in radians) as a function of confinement $\chi$ of tank-treading vesicles having different viscosity contrast ($\Lambda=0.5$, $1$, $4$, $8$). For $\Lambda > 1$, $\Psi$ varies non-monotonically with $\chi$. At higher confinement, all curves converge to the same value $\Psi = 0.23$ (radian). $\chi ^{*}=0.75$ is the degree of confinement when the channel half-height has the same size as the vesicle long axis.}
\end{figure}
\begin{figure}
\centering
\subfloat[]{\label{fig:pressure01}\includegraphics[angle=-90,width=0.3\textwidth]{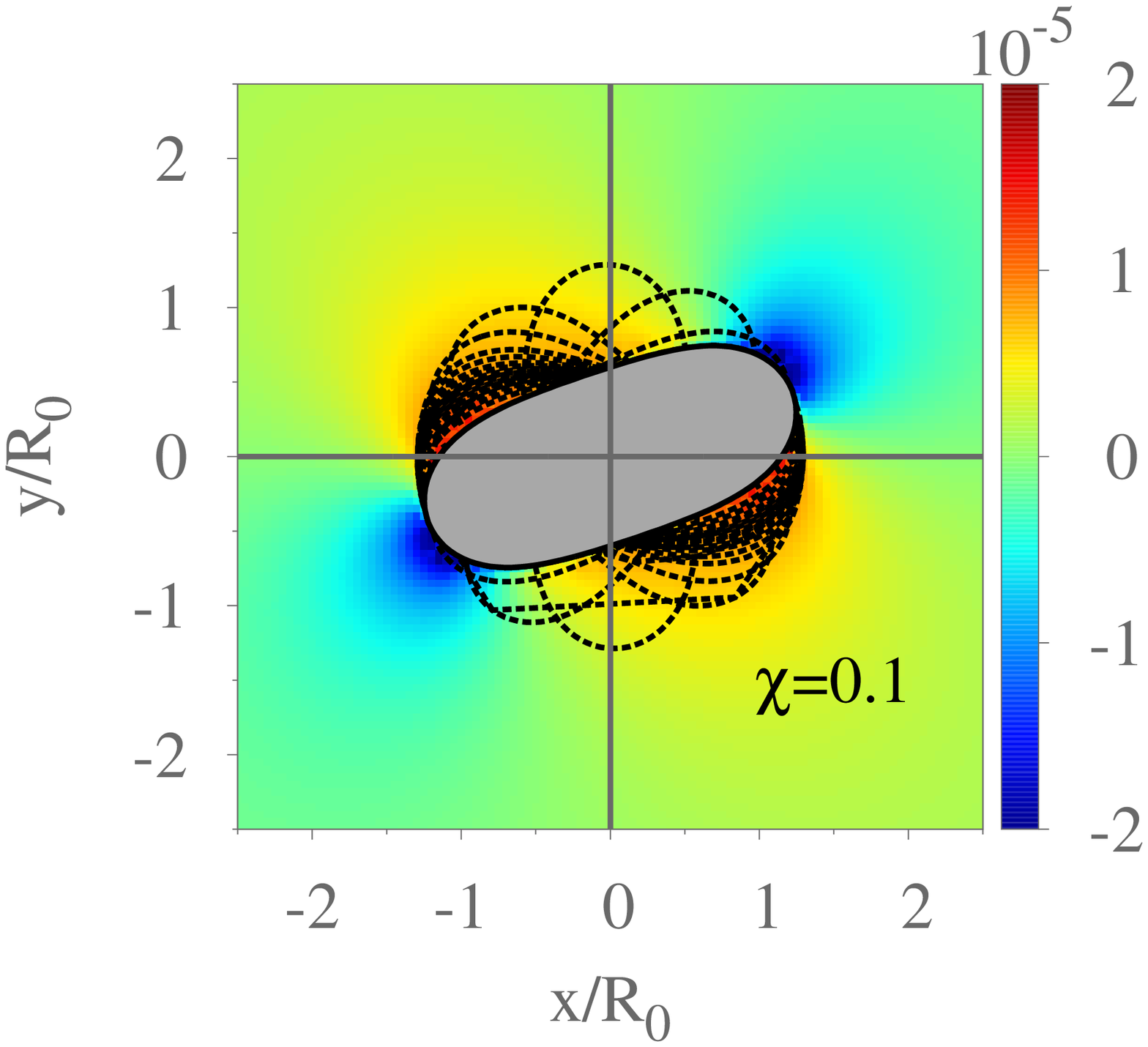}}
\qquad
\subfloat[]{\label{fig:pressure05}\includegraphics[angle=-90,width=0.3\textwidth]{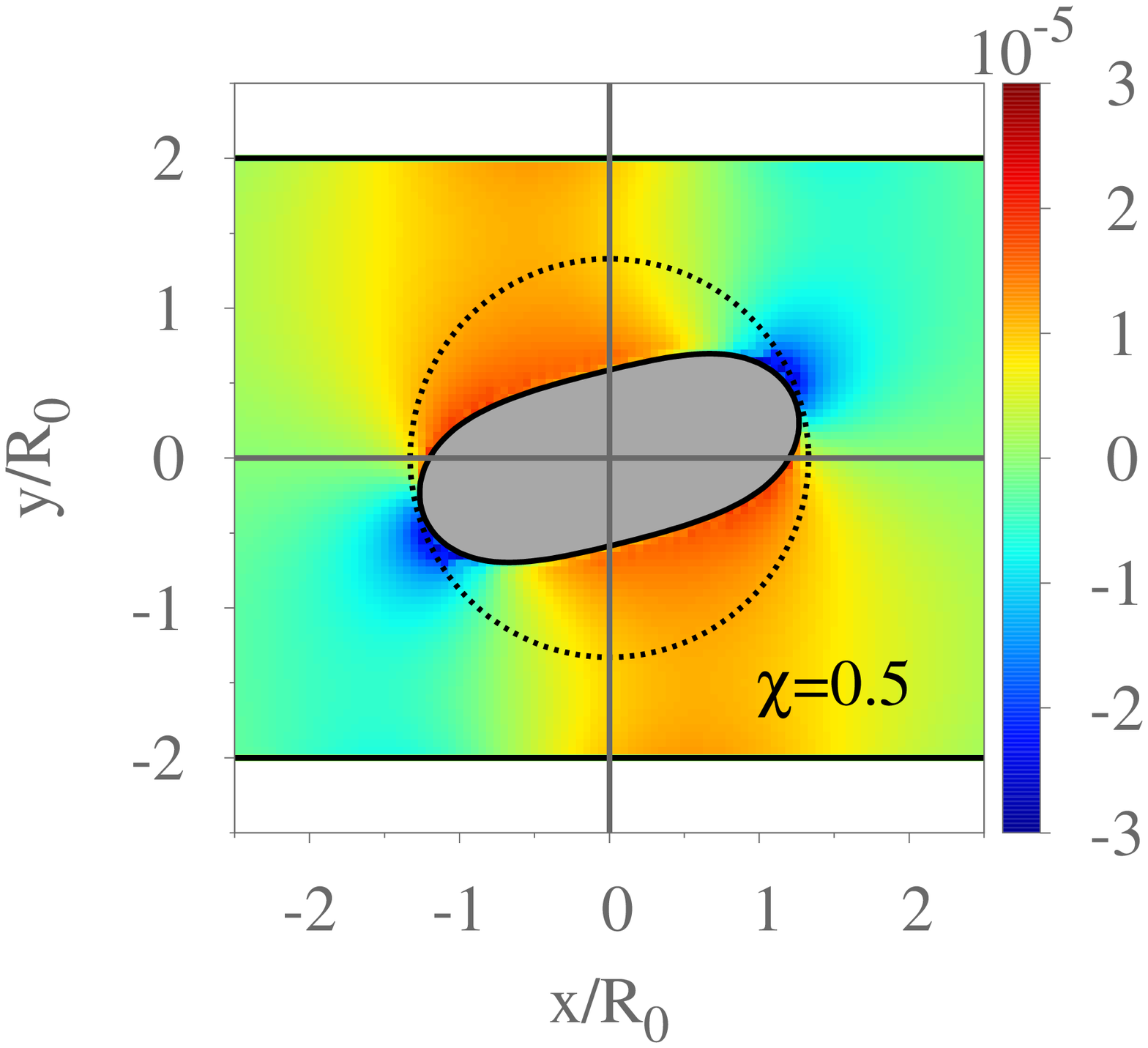}}
\qquad
\subfloat[]{\label{fig:pressure10}\includegraphics[angle=-90,width=0.3\textwidth]{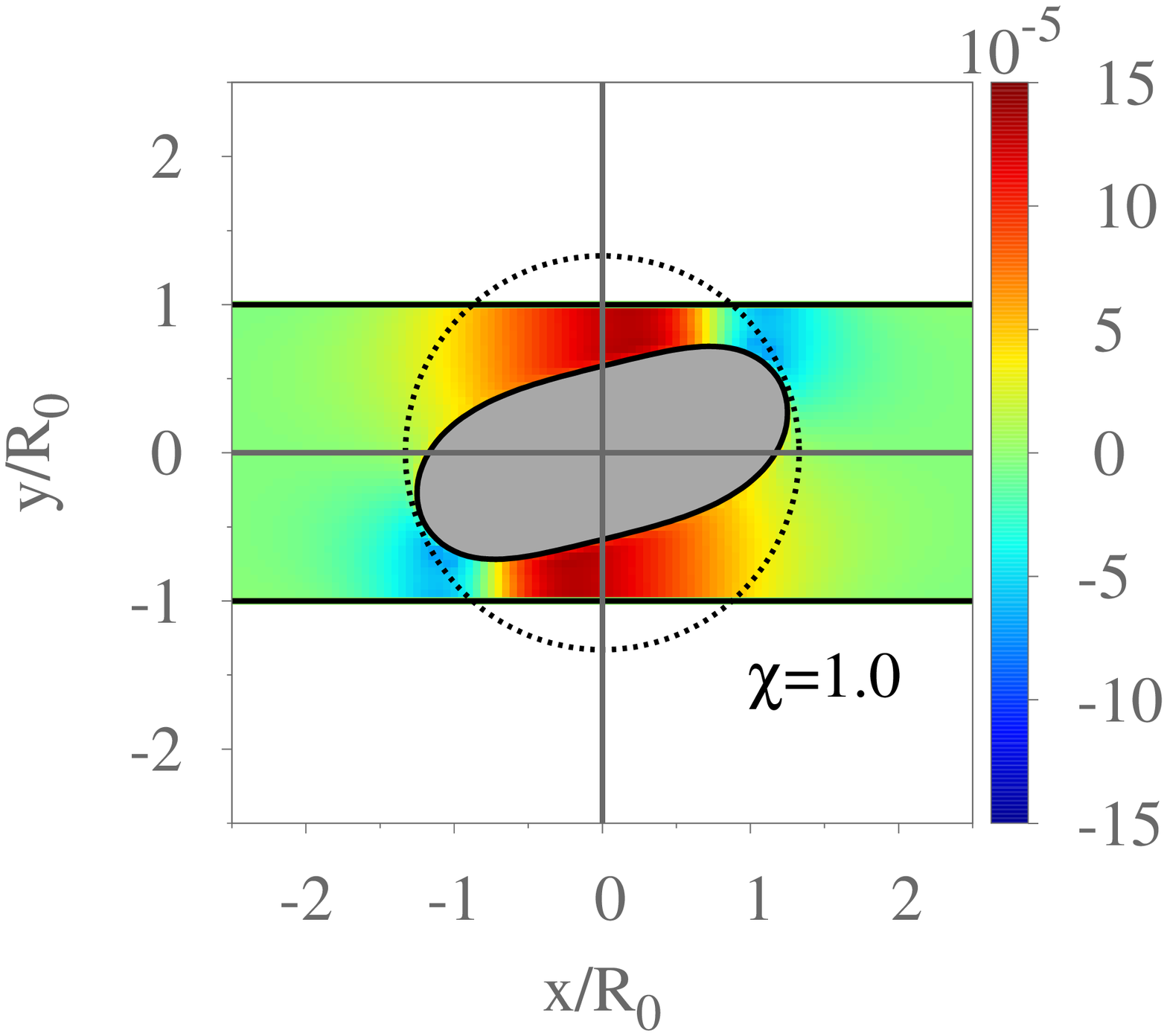}}
\caption{The pressure variation around the same viscous vesicle ($\Lambda =8$) subjected to shear flow at three different degrees of confinement: (a) lower $\chi=0.1$, (b) moderate $\chi=0.5$, and (c) higher $\chi=1.0$. The red-colored regions correspond to higher pressure, while the blue-colored ones correspond to lower pressure. In (a) the vesicle tumbles. The dashed lines are the vesicle snapshots taken at different times. In (b) and (c), the vesicle tank-treads. The circle (dotted line) shows the required space for tumbling.}
\label{fig:pressure}
\end{figure}

A step forward to answer the above question is to compare the variation of $\Psi$ with $\chi$ of tank-treading vesicles having different viscosity contrasts ($\Lambda = 0.5$, $1$, $4$, $8$). We notice that the vesicles behave differently depending on whether $\Lambda \leq 1$ or $\Lambda > 1$, see Fig.~{\ref{fig:angle_conf}}. For $\Lambda \leq 1$, the angle decreases monotonically with confinement $\chi$, it decreases rapidly when $\chi > 0.5$. For $\Lambda > 1$, the angle first increases up to a maximum and then decreases when the confinement becomes larger. For both cases ($\Lambda \leq 1$ and $\Lambda > 1$), the inclination angles converge to the same value ($\Psi = 0.23$ radian) at higher confinement ($\chi>1$). We observe the same non-monotonic variation of the angle with confinement for other values of $\Lambda > 1$ (not reported here). We are not aware of any work reporting this observation for vesicles or RBCs. However, for confined viscous droplets, the increase of $\Psi$ with $\chi$ has been observed by Janssen \textit{et al} numerically \cite{Janssen2008} and experimentally \cite{Janssen2010}, but no physical explanation for this behavior was provided. The authors rather used the angle variation with confinement to explain the droplet stability against break-up. Unlike viscous vesicles and RBCs, droplets are known to perform only the tank-treading motion, and therefore no dynamical state transition is observed for them. 
We explain the observed increase of $\Psi$ with $\chi$, for viscous vesicles, by analyzing the pressure field. Fig.~\ref{fig:pressure} shows the variation in the pressure around a sheared viscous vesicle ($\Lambda =8$) at different degrees of confinement ($\chi = 0.1$, $0.5$, $1.0$). The pressure values reported in Fig.~\ref{fig:pressure} are the relative deviation from the pressure of the external fluid in the absence of the vesicle. Regions with higher pressure are shown in red, those with lower pressure in blue. At lower confinement ($\chi = 0.1$, Fig.~\ref{fig:pressure}a), the rotational component of the imposed shear flow induces a viscous torque in clockwise direction responsible for the tumbling motion of the vesicle. Dashed closed lines in Fig.~\ref{fig:pressure}a represent the vesicle shapes (taken at equal time intervals) to show the tumbling motion. The vesicle rotates slowly and spends much time in the region of $-\pi/4 < \Psi < \pi/4$. By increasing confinement to a moderate value ($\chi = 0.5$, Fig.~\ref{fig:pressure}b), the pressure variations increase. Due to the pressure distribution around the vesicle (decreased pressure in the top right and bottom left quadrants and increased pressure in the top left and bottom right quadrants), a counterclockwise torque is induced. This pressure-induced torque competes with the torque due to the viscous stress mentioned above. At this confinement ($\chi=0.5$), the sum of the two torques vanishes. This results in a steady inclination of the vesicle, instead of tumbling and it explains the occurrence of the confinement-induced transition as observed in Figs.~\ref{fig:angle_visco}, \ref{fig:transition_conf} and \ref{fig:bifurcation_conf}. Increasing the confinement further amplifies the pressure-induced torque and consequently increases the inclination angle. This explains the increase of the angle with confinement observed in Fig.~\ref{fig:bifurcation_conf} and \ref{fig:angle_conf}. At larger degrees of confinement ($\chi = 1.0$, Fig.~\ref{fig:pressure}c) the vesicle approaches the wall. In this situation, the gap between the membrane and the wall becomes thin. However, the vesicle never touches the walls due to lubrication forces that induce a clockwise torque pushing the vesicle away from the walls. This coincides with observing a negative local wall shear stress (like in \cite{Kaoui2011a}). At the largest confinement studied ($\chi = 1.25$), the steady angle is $\Psi = 0.23$ (radian), regardless of the viscosity contrast (see Fig.~{\ref{fig:angle_conf}}). At this degree of confinement, the angle is mainly governed by the geometry (channel height and vesicle shape), and the other parameters, especially the viscosity contrast, do not play a significant role. This is caused by the balance of viscous-, pressure- and lubrication-induced torque. 
\begin{figure}[b]
\captionsetup{type=figure}
\centering
\subfloat[]{\label{fig:angular_a}\includegraphics[angle=-90,width=0.24\textwidth]{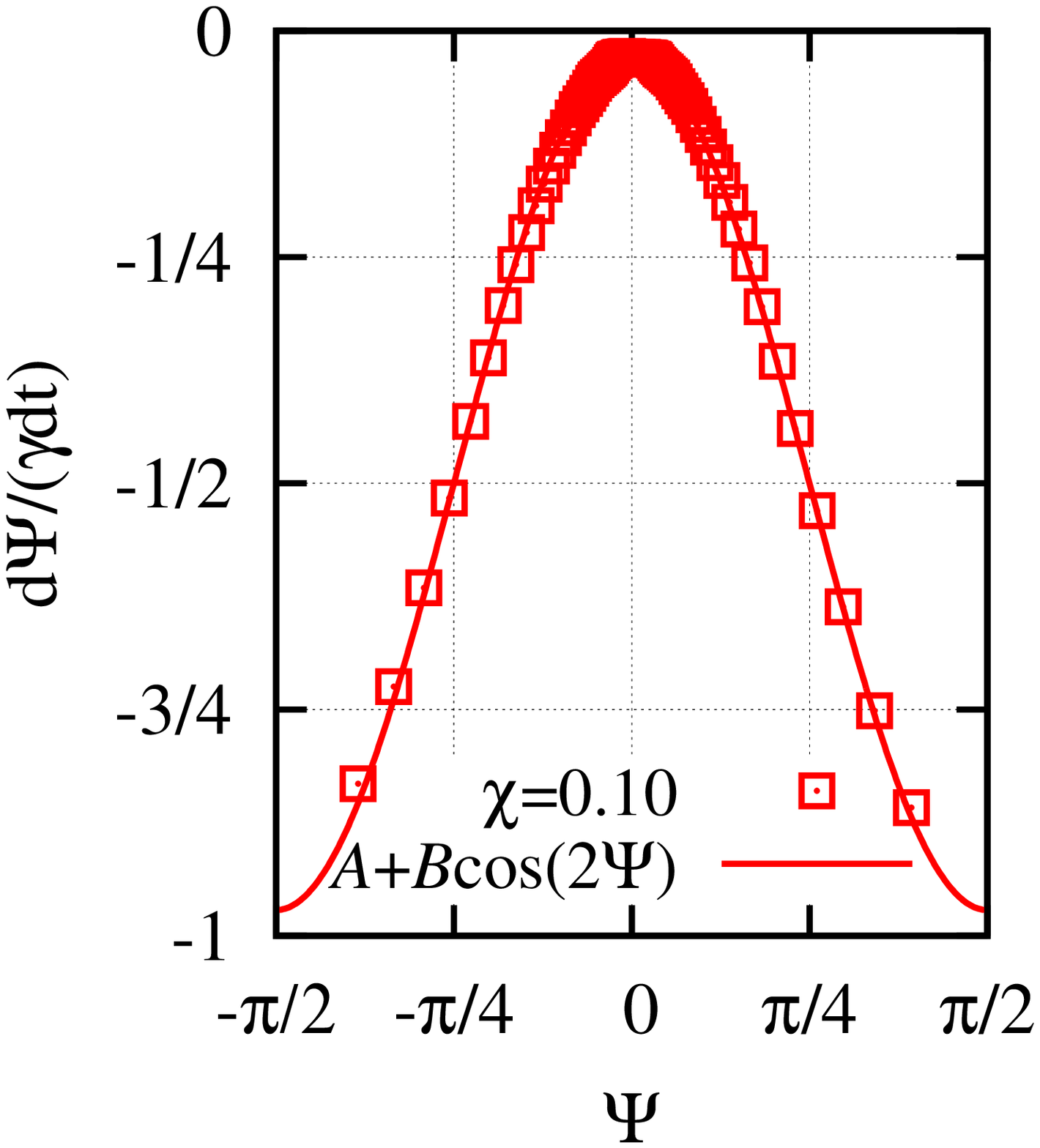}}
~
\subfloat[]{\label{fig:angular_b}\includegraphics[angle=-90,width=0.2125\textwidth]{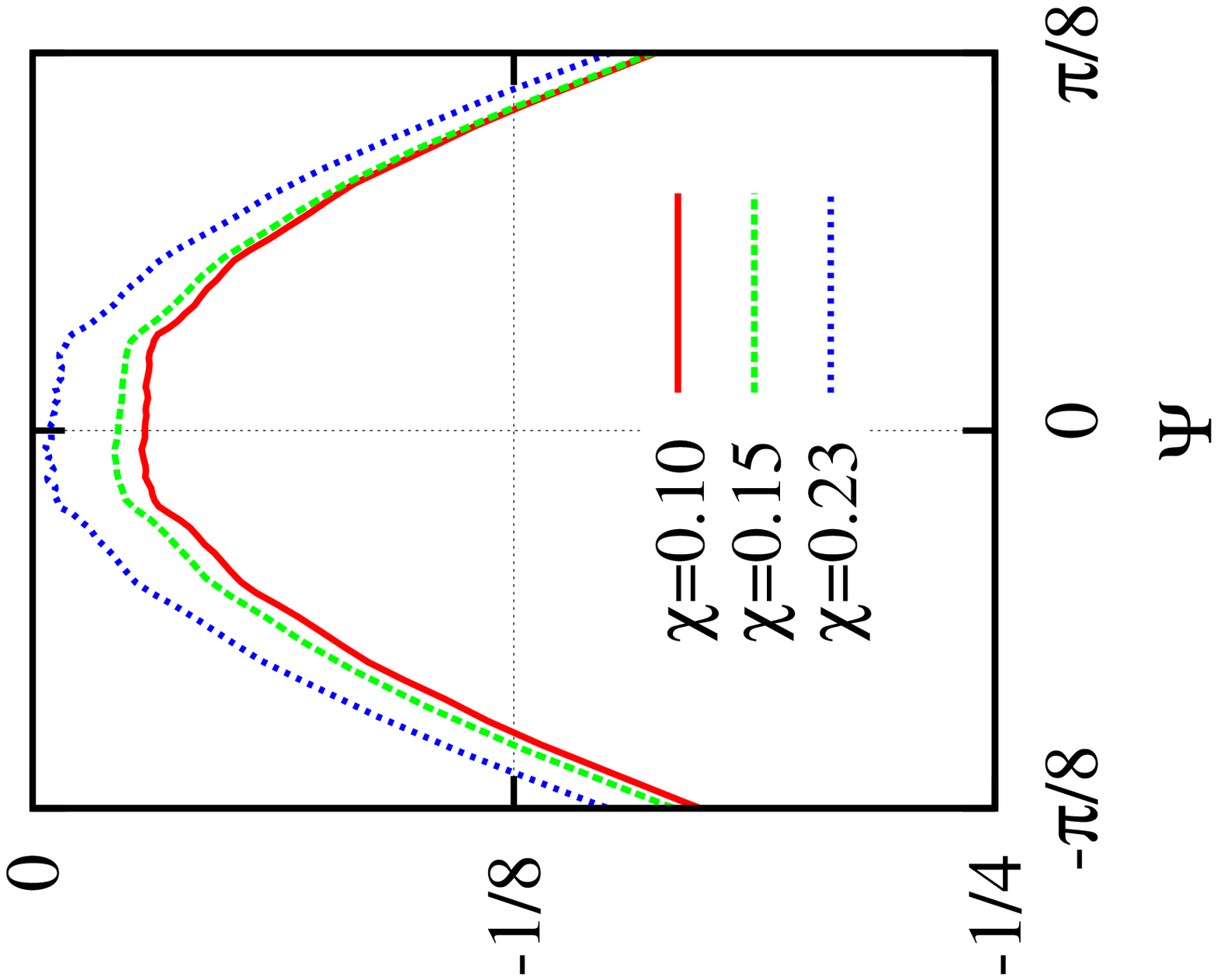}}
\caption{The angular velocity ${\rm d}\Psi/(\gamma {\rm d}t)$ versus the instantaneous inclination angle $\Psi$ during the tumbling motion of a viscous vesicle ($\Lambda =8$). (a) the data-set for $\chi =0.1$ (square symbols) is fitted perfectly with the function $A+B\cos(2\Psi)$ (solid line), where $A=1/2$ and $B=0.47$. (b) the angular velocity versus the angle for different degrees of confinement ($\chi =0.10$, $0.15$, $0.23$). At higher $\chi$, the angular velocity approaches zero.}
\label{fig:angular}
\end{figure}

After explaining the mechanism behind the behavior of the vesicle angle with confinement, now we analyze the effect of confinement on the way the vesicle tumbles. In addition to the tumbling frequency (shown in Fig.~\ref{fig:bifurcation_conf}) we analyze the phase portrait, where the normalized angular velocity $\text{d}\Psi/\gamma \text{d}t$ is plotted as a function of the instantaneous inclination angle $\Psi(t)$, see Fig.~\ref{fig:angular}. The angular velocity is negative since the vesicle performs a clockwise tumbling motion. The value ${\rm d}\Psi/{\rm d}t = -\gamma/2$ corresponds to the angular velocity of a rigid cylinder rotating in unbounded shear flow \cite{Cox1968}. It has the same value independent of $\Psi$. For deflated vesicles (here $\Delta = 0.8$), ${\rm d}\Psi/{\rm d}t > -\gamma/2$ when the angle is between $-\pi /4$ and $\pi /4$, thus, the vesicle is slower and spends more time in that angular region. For the remaining range of angles, ${\rm d}\Psi/{\rm d}t < -\gamma/2$, and the vesicle rotates faster. The solid line in Fig.~\ref{fig:angular}a is a fit of the numerical data-set (square symbols) with the function
\begin{equation}
\frac{1}{\gamma}\frac{\text{d} \Psi}{\text{d}t} = A + B \cos (2\Psi)
\label{eq:KS}
\end{equation}
introduced by Jeffery to describe the rotation of rigid ellipsoidal particles in unbounded shear flow \cite{Jeffery1922}. Here, $A = -1/2$ and $B$ is a function of the particle aspect ratio (ratio between the long and the short axes in the shearing plane). Eq.~\ref{eq:KS} has since been extended by Keller and Skalak \cite{Keller1982} to take into account the motion of fluid-filled particles having a liquid membrane (such as RBCs or vesicles). In this case, $B$ depends on the viscosity contrast $\Lambda$. The theory of Keller and Skalak (KS) predicts the dynamics of a vesicle in unbounded shear flow, depending on the value of $-A/B$. For $-A/B > 1$, the vesicle tumbles, whereas for $0 < -A/B < 1$, the vesicle tank-treads. It was found experimentally that, for larger deformation (${\rm Ca} \gg 1$), the terms $A$ and $B$ deviate from their respective expressions given by the theory of KS \cite{Mader2006}. In the present paper, ${\rm Ca}=0.5$ is sufficiently small. Therefore, $A$ is considered to be a constant and equal to $- 1 / 2$. Only $B$ is assumed to be a function of $\Lambda$ and $\chi$, and used as a fitting parameter. Obviously, Eq.~(\ref{eq:KS}) accurately describes the evolution of the inclination angle for $\chi = 0.1$, see Fig.~\ref{fig:angular}a. The angular velocity $\text{d}\Psi/(\gamma \text{d}t)$ is plotted as a function of $\Psi$ for three values of confinement ($\chi = 0.10$, $0.15$, $0.23$) in Fig.~\ref{fig:angular}b. For convenience, only the region $[-\pi/8,\pi /8] \times [-1/4,0]$ is shown. By increasing $\chi$, the maximum of the angular velocity approaches zero. This means that the motion of the vesicle slows down in the region $[-\pi/4,\pi /4]$ where it spends most of the time. As a result, the tumbling is delayed. Presumably, a further increase of $\chi$ ($\chi > 0.23$) would lead to a positive value of $\text{d}\Psi/\text{d}t$ and therefore to an unstable counterclockwise rotation. Instead, we observe the transition to the tank-treading motion.
\begin{figure}[t]
\resizebox{\columnwidth}{!}{\includegraphics[angle=-90]{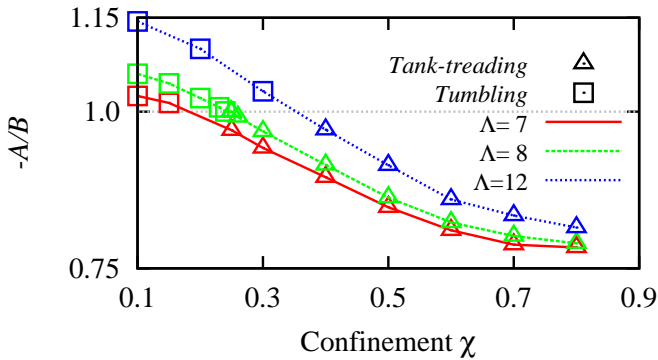}}
\caption{\label{fig:ks}
Variation of the Keller and Skalak parameter $-A/B$ with confinement $\chi$ for different viscosity contrast ($\Lambda=7$, $8$, $12$). For each $\Lambda$, the transition from tumbling to tank-treading occurs exactly when $-A/B=1$. The corresponding critical value of $\chi$, for each $\Lambda$, agrees with the one reported in Fig.~\ref{fig:transition_conf}.}
\end{figure}

By fitting the data in Fig.~\ref{fig:angular}b with Eq.~(\ref{eq:KS}) we find that $B$ is an increasing function of confinement $\chi$. In Fig.~\ref{fig:ks}, we report $-A/B$ versus $\chi$, for $\Lambda=7$, $8$, $12$. Square symbols correspond to the values obtained when the vesicle tumbles (using a fit as described above). For these points, we observe $-A/B > 1$. The value of $-A/B$ decreases until reaching unity at a critical confinement $\chi _C$ where the transition from tumbling to tank-treading occurs. The corresponding values of $\chi _C$ for each $\Lambda$ are in good agreement with the phase-diagram in Fig.~\ref{fig:transition_conf}. The triangles in Fig.~\ref{fig:ks} denote the values of $-A/B$ computed for tank-treading vesicles via the equation $-A/B = \cos (2\Psi)$ which is Eq.~(\ref{eq:KS}) for a steady inclination angle (${\rm d}\Psi/{\rm d} t=0$). For these points, $0 < -A/B < 1$ holds. As mentioned above, in the KS theory, it is sufficient to study the value of $-A/B$ in order to predict the vesicle dynamical state in unbounded shear flow. Interestingly, this also applies here to catch the transition, although the system is confined, while the KS theory does not include confinement effects. We emphasize again that here the transition is induced by varying $\chi$ alone. The viscosity contrast, the swelling degree, and the capillary number are all kept constant. Thus, the transition from tumbling to tank-treading due to confinement can also be explained based on the theory of KS. The transition takes place when $-A/B=1$. Above, we used values of $B$ obtained from our simulations. Even though further theoretical study is needed to derive a closed analytical expression for the $\chi$-dependence of $B$, we show that $B$ is an increasing function of $\chi$.
\section{Conclusions}
Using computer simulations we show that by only squeezing a viscous vesicle one can induce the dynamical transition from solid-like motion (tumbling) to liquid-like motion (tank-treading). Moreover, we find that confinement slows down the tumbling motion and increases the inclination angle of tank-treading vesicles with $\Lambda > 1$. The confinement-induced transition and associated vesicle dynamical behavior result from the competition between viscous- and pressure-induced torques exerted upon the vesicle. At larger degrees of confinement, the lubrication-induced torque becomes dominant and pushes the vesicle to align with the walls. The dynamics of vesicles are well described in the limit of the low deformation regime by the theory of Keller and Skalak \cite{Keller1982}. However, it is limited to unbounded shear flow. Here, we predict that the term $B$ in this theory would be a function of confinement. The tumbling-to-tank-treading transition is also observed for RBCs. It is usually attributed to the increase in the shear rate $\gamma$, for instance, during the displacement of RBCs from large arteries (where $\gamma$ is lower) to the branches (where $\gamma$ is higher). A RBC also stops tumbling when it is squeezed in a capillary, because of the lack of available space. It is necessary for RBCs to switch to tank-treading mode in order to pass through narrower capillaries. Based on our findings, a RBC may be expected to undergo this transition even at constant shear rate and at lower confinement ($\chi < 0.5$), compared to $\chi > 1$ encountered in capillaries. The effect of confinement can be seen as an extra contribution (besides the change in shear rate) to the transition from tumbling to tank-treading of RBCs. This transition is found to be a contributing factor for shear thinning of blood \cite{Forsyth2011}. Therefore, since we could induce this dynamical transition by confinement we expect that this will also lead to shear thinning for a suspension of viscous vesicles or RBCs. Under these circumstances, the walls play the role of inducing the transition; in contrast to the Fahraeus-Lindqvist effect where the walls induce cross-stream migration of RBCs. We expect that our study will stimulate further studies, similar to Ref.~\cite{Peyla2011}, for confined suspensions of deformable micro-particles in order to gain more insight into the link between confinement effects, dynamical transition and the effective viscosity. 
\begin{acknowledgments}
We thank NWO/STW (VIDI grant 10787 of Jens Harting) and the TU/e High Potential Research Program for financial support.
\end{acknowledgments}
\end{document}